\documentclass{article}
\usepackage{spconf,amsmath,graphicx,amssymb}
\usepackage[export]{adjustbox}
\usepackage{subfig}
\usepackage{color}
\usepackage{enumitem}
\PassOptionsToPackage{hyphens}{url}\usepackage{hyperref}
\newcommand{\vx}{\mathbf{x}}
\newcommand{\ve}{\boldsymbol{\omega}}
\newcommand{\vh}{\mathbf{h}}
\newcommand{\vb}{\mathbf{b}}
\newcommand{\vv}{\mathbf{v}}
\newcommand{\vc}{\mathbf{c}}
\newcommand{\vw}{\mathbf{w}}
\newcommand{\vW}{\mathbf{W}}

\newcommand{\ie}{\textit{i.e.}, }
\def\x{{\mathbf x}}

\hypersetup{colorlinks=true}

\title{Attention-Based Models for Text-Dependent Speaker Verification}

\twoauthors
{F A Rezaur Rahman Chowdhury\sthanks{The author did this work during his intern at Google.}}
{Washington State University\\
\normalsize
\href{mailto:fchowdhu@eecs.wsu.edu}{\nolinkurl{fchowdhu@eecs.wsu.edu}}
}
{Quan Wang, Ignacio Lopez Moreno, Li Wan}
{Google Inc., USA\\
\normalsize
\{\href{mailto:quanw@google.com}{\nolinkurl{quanw}},
\href{mailto:elnota@google.com}{\nolinkurl{elnota}},
\href{mailto:liwan@google.com}{\nolinkurl{liwan}}\}
{\tt @google.com}
}

\begin{document}
\ninept
\maketitle
\begin{abstract}
Attention-based models have recently shown great performance on a range of tasks,
such as speech recognition, machine translation, and image captioning
due to their ability to summarize relevant information that expands through the
entire length of an input sequence.
In this paper, we analyze the usage of attention mechanisms to the problem of
sequence summarization in our end-to-end text-dependent speaker recognition system.
We explore different topologies and their variants of the attention layer,
and compare different pooling methods on the attention weights.
Ultimately, we show that attention-based models can improves the
Equal Error Rate (EER) of our speaker verification system by relatively 14\% compared to our
non-attention LSTM baseline model.
\end{abstract}
\section{Introduction}
\label{sec:intro}

Speaker verification (SV) is the process of verifying, based on a set of reference enrollment utterances, whether an verification utterance belongs to a known speaker.
One subtask of SV is global password text-dependent speaker verification (TD-SV), which refers to the set of problems for which the transcripts of reference enrollment and verification utterances are constrained to a specific phrase.
In this study, we focus on ``OK Google" and ``Hey Google" global passwords, as
they relate to the Voice Match feature of Google Home \cite{multiuser,voicematch}.

\textit{I-vector} \cite{dehak2011front} based systems in combination with verification back-ends such as Probabilistic Linear Discriminant Analysis (PLDA) \cite{garcia2011analysis} have been the dominating paradigm of SV in previous years. More recently, with the rising of deep learning \cite{lecun2015deep} in various machine learning applications, more efforts have been focusing on using neural networks for speaker verification. Currently, the most promising approaches are
end-to-end integrated architectures that simulate the enrollment-verification two-stage process
during training.

For example, in \cite{rohdin2017end} the authors propose architectures that resemble the components of an i-vector + PLDA system. Such architecture allowed to bootstrap the network parameters from pretrained i-vector and PLDA models for a better performance.
However, such initialization stage also constrained the type of network architectures that could be used --- only Deep Neural Networks (DNN) can be initialized from classical i-vector and PLDA models.
In \cite{heigold2016end}, we have shown that Long Short-Term Memory (LSTM) networks \cite{hochreiter1997long} can achieve better performance than DNNs for integrated end-to-end architectures in TD-SV scenarios.

However, one challenge in our architecture introduced in \cite{heigold2016end} is that, silence and background noise are not being well captured. Though our speaker verification runs on a short 800ms window that is segmented by the keyword detector \cite{chen2014small,prabhavalkar2015automatic},
the phonemes are usually surrounded by frames of silence and background noise. Ideally, the speaker embedding should be built only using the frames corresponding to phonemes.
Thus, we propose to use an attention layer \cite{chorowski2015attention,luong2015effective,xu2015show} as a soft mechanism to emphasize the most relevant elements of the input sequence.

This paper is organized as follows. In Sec. \ref{sec:baseline}, we first briefly review
our LSTM-based d-vector baseline approach trained with the end-to-end
architecture \cite{heigold2016end}.
In Sec. \ref{sec:attention}, we introduce how we add the attention mechanism to
our baseline architecture, covering different scoring functions, layer variants,
and weights pooling methods. In Sec. \ref{sec:exp} we setup experiments to compare
attention-based models against our baseline model, and present the EER results
on our testing set. Conclusions are made in Sec. \ref{sec:conclusions}.

\section{Baseline Architecture}
\label{sec:baseline}

\begin{figure}
  \centering
    \includegraphics[width=0.35\textwidth]{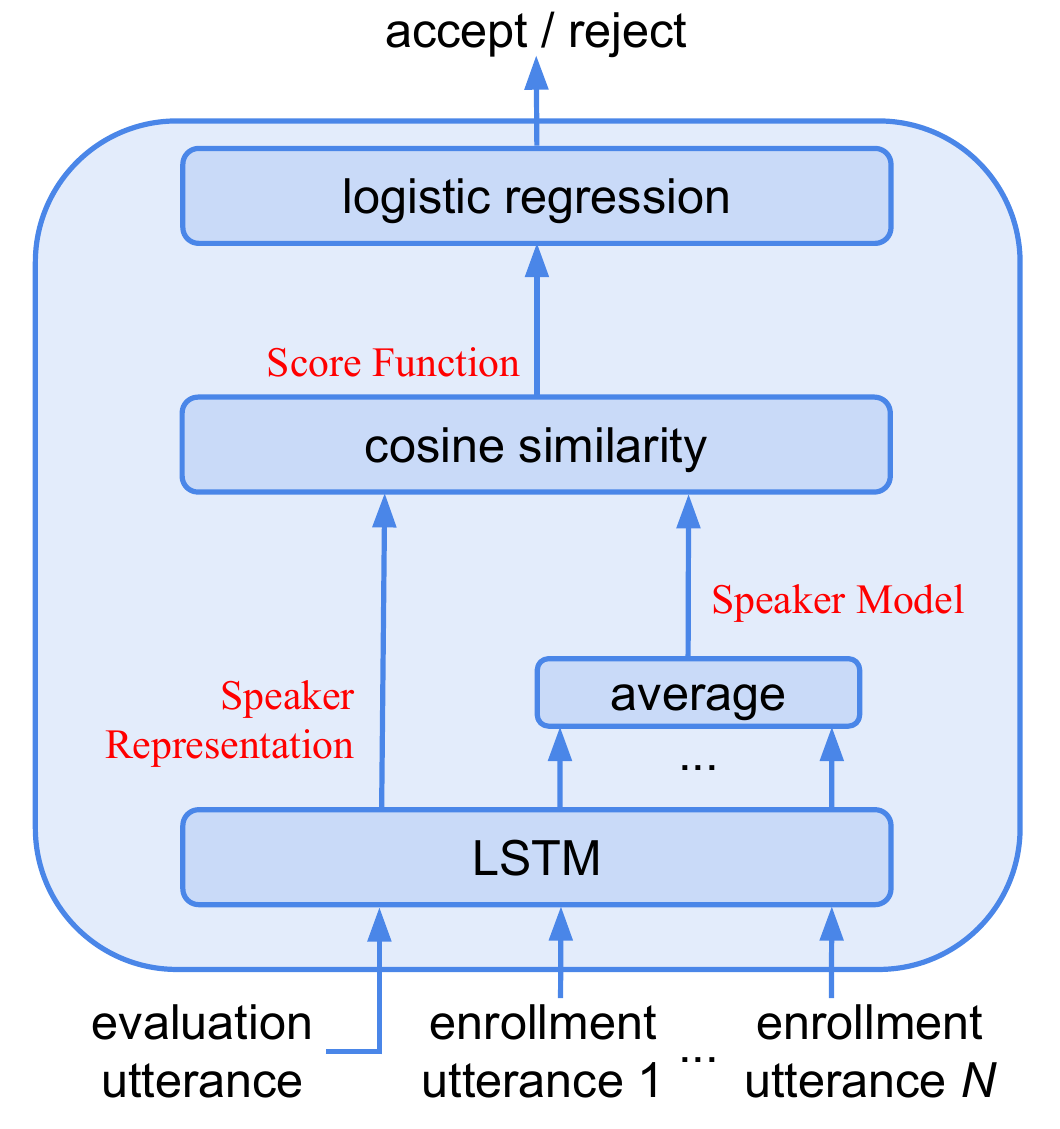}
  \caption{Our baseline end-to-end training architecture as introduced in \cite{heigold2016end}.}
  \label{fig:e2e}
\end{figure}

Our end-to-end training architecture \cite{heigold2016end} is described in Fig. \ref{fig:e2e}.
For each training step, a tuple of one evaluation utterance $\vx_{j\sim}$ and $N$
enrollment utterances $\vx_{kn}$ (for $n=1,\cdots,N$) is fed into our LSTM network:
$\{\vx_{j\sim},(\vx_{k1},\cdots, \vx_{kN})\}$,
where $\vx$ represents the features (log-mel-filterbank energies) from a fixed-length segment,
$j$ and $k$ represent the speakers of the utterances, and $j$ may or may not equal $k$.
The tuple includes a single utterance from speaker $j$, and $N$ different utterance from speaker $k$.
We call a tuple positive if $\x_{j\sim}$ and the $N$ enrollment utterances are
from the same speaker, \ie $j=k$, and negative otherwise.
We generate positive and negative tuples alternatively.

For each utterance, let the output of the LSTM's last layer at frame $t$ be a fixed dimensional vector $\vh_t$, where $1 \leq t \leq T$. We take the \textit{last frame output} as the d-vector $\ve = \vh_T $ (Fig. \ref{fig:baseline}), and build a new tuple:
$\{\ve_{j\sim}, (\ve_{k1},\cdots, \ve_{kN})\}$.
The centroid of
tuple $(\ve_{k1},\cdots, \ve_{kN})$ represents the voiceprint built from $N$ utterances,
and is defined as follows:
\begin{equation}
  \label{eqn:centroid}
  \vc_k=\mathrm{E}_n [ \ve_{kn} ]=\frac{1}{N}\sum_{n=1}^N \dfrac{\ve_{kn}}{\mathbin{\|}\ve_{kn}\mathbin{\|}_2}.
\end{equation}
The similarity is defined using the cosine similarity function:
\begin{equation}
\label{eqn:similarity_old}
  s=w\cdot \cos(\ve_{j\sim}, \vc_{k})+b,
\end{equation}
with learnable
$w$ and $b$.  The tuple-based end-to-end loss is finally defined as:
\begin{equation}
  L_{\mathrm{T}}(\ve_{j\sim}, \vc_k)=\delta(j,k)\sigma(s)+ \Big( 1-\delta(j,k) \Big)\Big(1- \sigma(s)\Big) .
\end{equation}
Here $\sigma(x)=1/(1+e^{-x})$ is the standard sigmoid function and
$\delta(j,k)$ equals $1$ if $j=k$, otherwise equals to $0$.
The end-to-end loss function encourages a larger value of $s$ when $k=j$, and a smaller
value of $s$ when $k\neq j$. Consider the update for both positive and
negative tuples --- this loss function
is very similar to the triplet loss in FaceNet~\cite{SchroffKP15}.

\section{Attention-based model}
\label{sec:attention}

\subsection{Basic attention layer}
\label{sec:basic}

\begin{figure}
\centering
\subfloat[]{
  \includegraphics[width=.31\textwidth,valign=c]{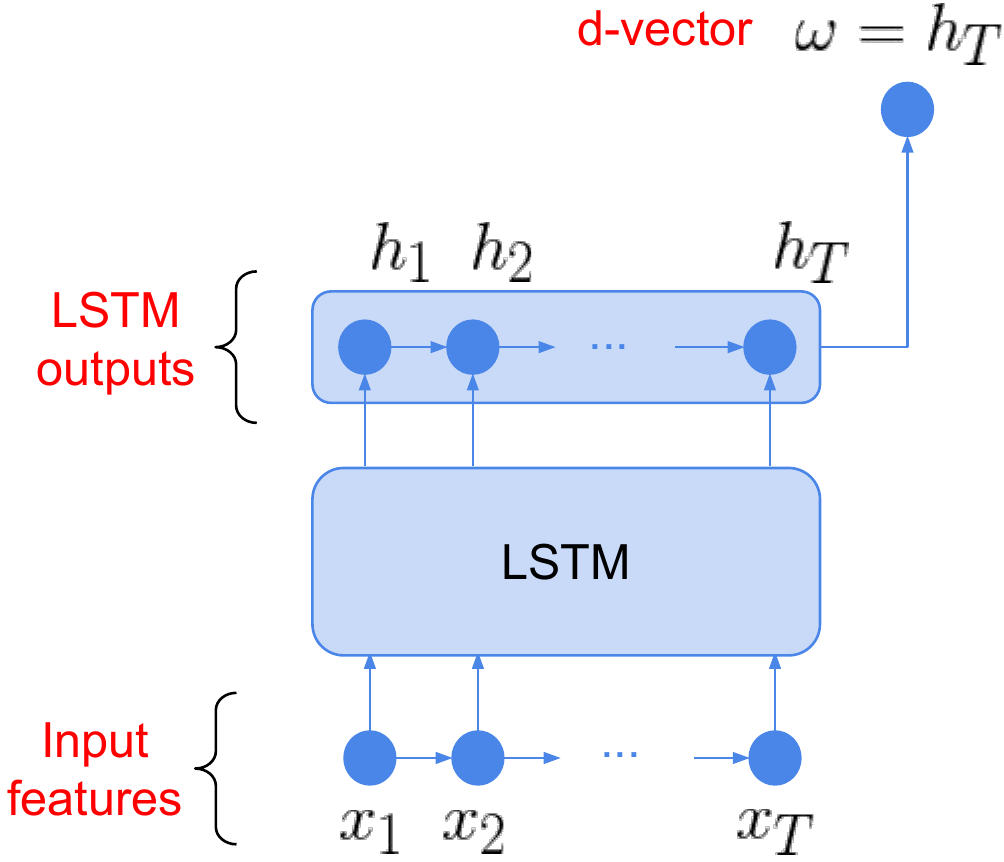}
  \label{fig:baseline}
}
\\
\subfloat[]{
  \includegraphics[width=.31\textwidth,valign=c]{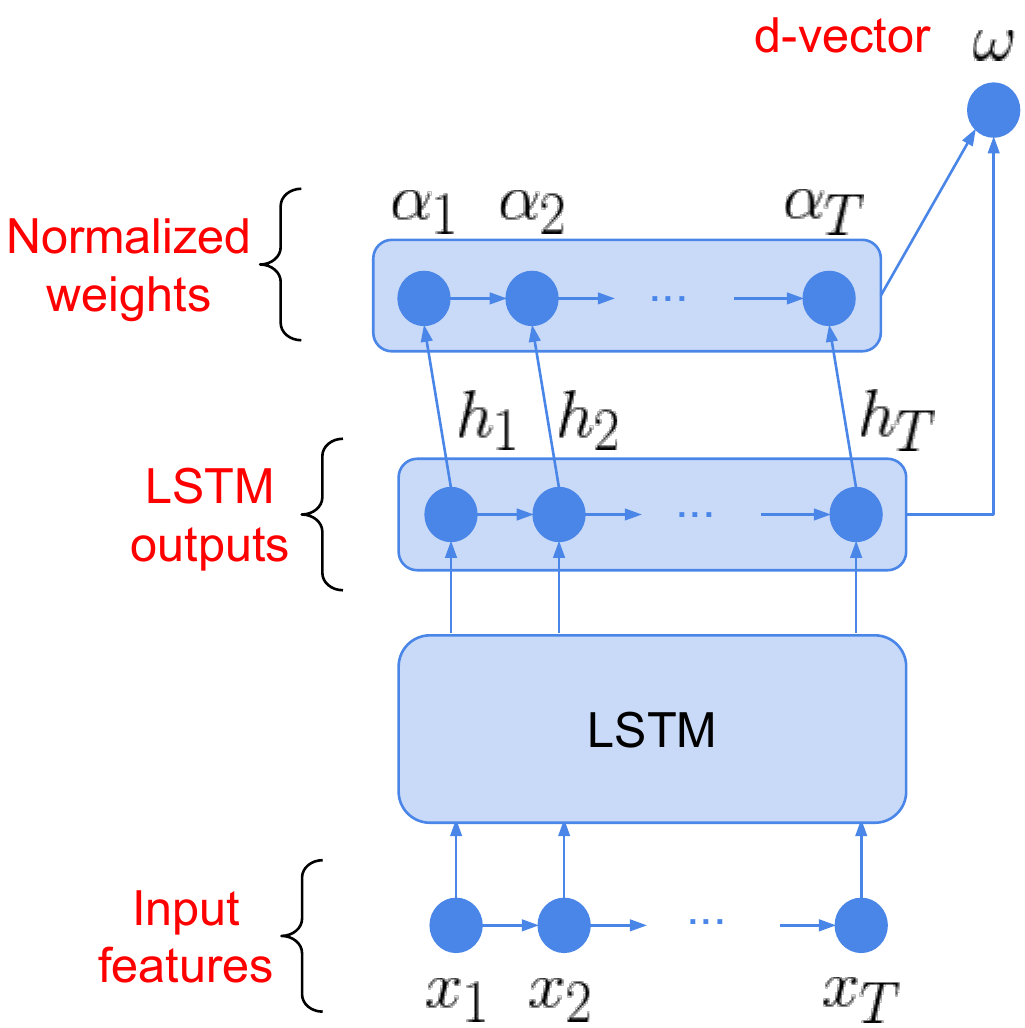}
  \label{fig:attention_layer}
}
\caption{(a) LSTM-based d-vector baseline \cite{heigold2016end}.
(b) Basic attention layer.}
\label{fig:basic_attention}
\end{figure}

In our baseline end-to-end training, we directly take the last frame output as d-vector $\ve = \vh_T $.
Alternatively, we could learn a scalar score $e_t \in \mathbb{R}$ for the LSTM output $\vh_t$ at each frame $t$:
\begin{equation}
\label{eq:score}
e_t=f(\vh_t) ,
\qquad t=1,\cdots,T .
\end{equation}
Then we can compute the normalized weights $\alpha_t \in [0,1]$ using these scores:
\begin{equation}
\label{eq:weight}
\alpha_t = \dfrac{\exp(e_t)}{\sum_{i=1}^{T}\exp(e_i)} ,
\end{equation}
such that
$\sum_{t=1}^{T}\alpha_t = 1$.
And finally, as shown in Fig. \ref{fig:attention_layer}, we form the d-vector $\ve$ as the weighted average of the LSTM outputs
at all frames:
\begin{equation}
\label{eq:omega}
\ve = \sum_{t=1}^{T}\alpha_t \vh_t .
\end{equation}

\subsection{Scoring functions}
\label{sec:scoring}
By using different scoring functions $f(\cdot)$ in Eq. (\ref{eq:score}),
we get different attention layers:
\begin{itemize}[itemsep=1mm,topsep=1mm]
  \item Bias-only attention, where $b_t$ is a scalar. Note this attention does not depend on the LSTM output $\vh_t$.
  \begin{equation}
  e_t = f_\mathrm{BO}(\vh_t) = b_t .
  \end{equation}

  \item Linear attention, where $\vw_t$ is an $m$-dimensional vector, and $b_t$ is a scalar.
  \begin{equation}
  e_t = f_\mathrm{L}(\vh_t) = \vw_t^T \vh_t + b_t .
  \end{equation}

  \item Shared-parameter linear attention, where the $m$-dimensional vector $\vw$ and scalar $b$ are the same for all frames.
  \begin{equation}
  e_t = f_\mathrm{SL}(\vh_t) = \vw^T \vh_t + b .
  \end{equation}

  \item Non-linear attention, where $\vW_t$ is an $m' \times m$ matrix, $\vb_t$ and $\vv_t$ are $m'$-dimensional vectors.
  The dimension $m'$ can be tuned on a development dataset.
  \begin{equation}
  e_t = f_\mathrm{NL}(\vh_t) = \vv_t^T \tanh ( \vW_t \vh_t + \vb_t ) .
  \end{equation}

  \item Shared-parameter non-linear attention, where the same $\vW$, $\vb$ and $\vv$ are used for all frames.
  \begin{equation}
  e_t = f_\mathrm{SNL}(\vh_t) = \vv^T \tanh ( \vW \vh_t + \vb ) .
  \end{equation}

\end{itemize}

In all the above scoring functions, all the parameters are trainable within the end-to-end
architecture \cite{heigold2016end}.

\subsection{Attention layer variants}
\label{sec:variant}

Apart from the basic attention layer described in Sec. \ref{sec:basic}, here we introduce two variants: cross-layer attention, and
divided-layer attention.

\begin{figure}
\centering
\subfloat[]{
  \includegraphics[width=.45\textwidth,valign=c]{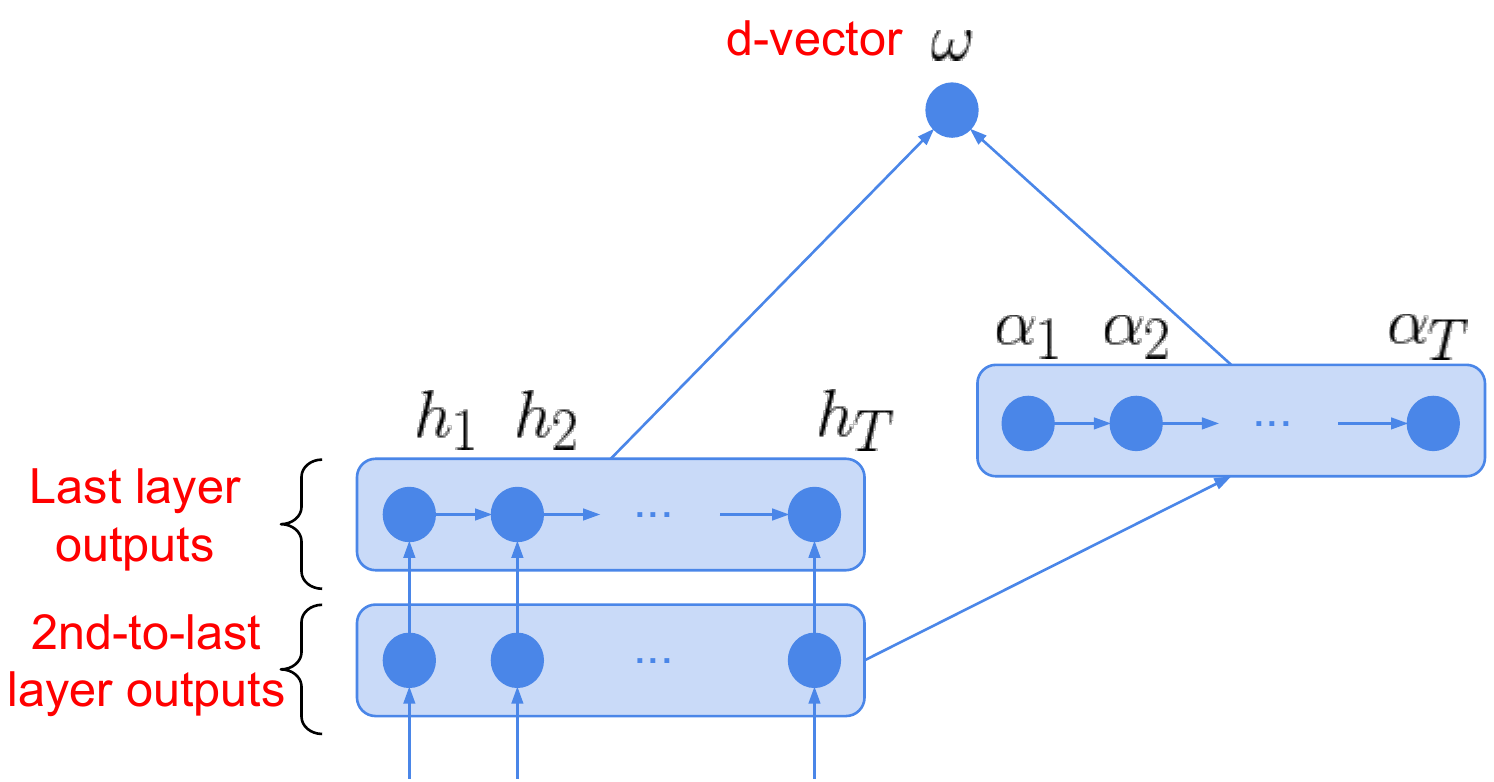}
  \label{fig:variantion1}
}\\
\subfloat[]{
  \includegraphics[width=.35\textwidth,valign=c]{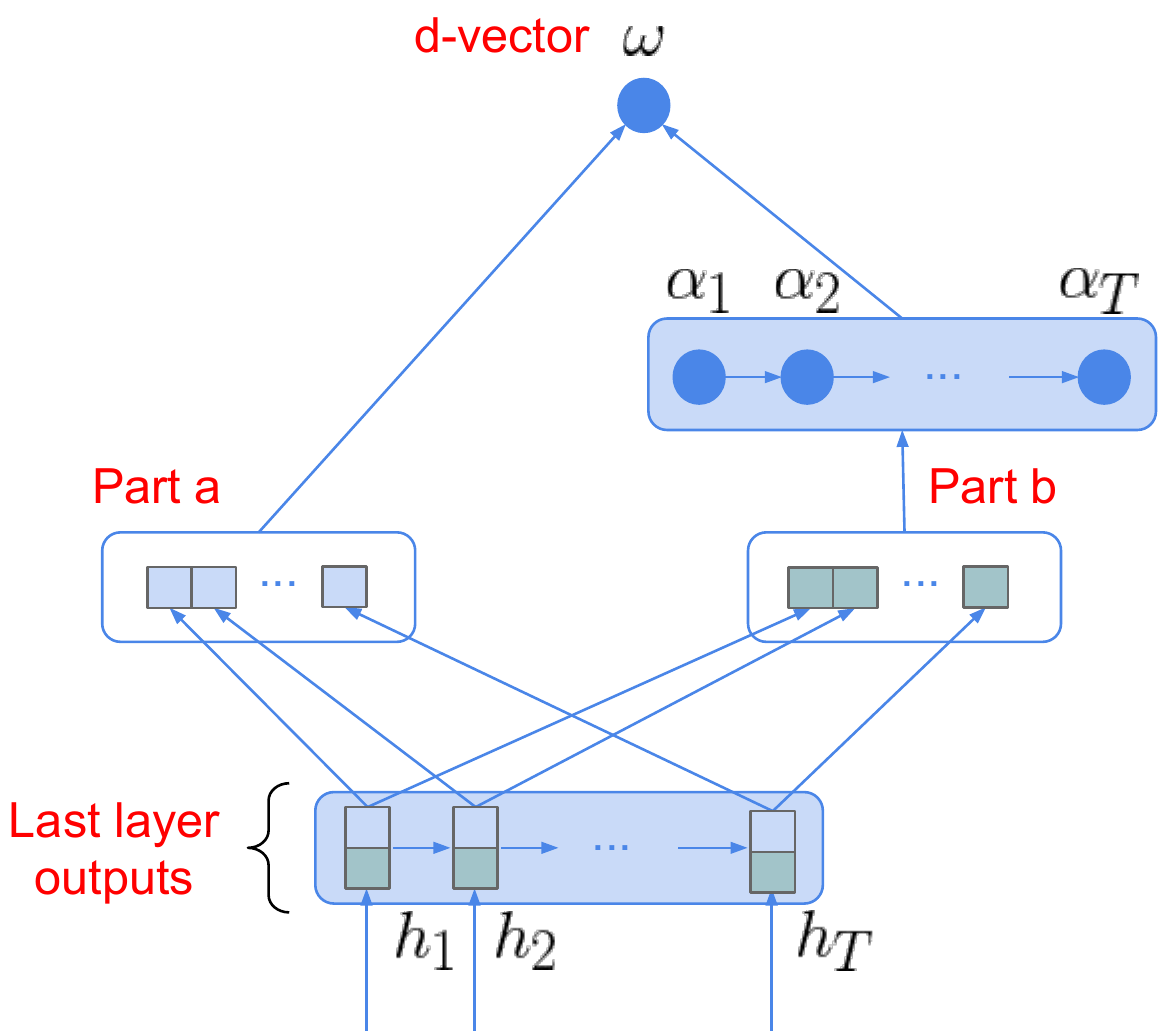}
  \label{fig:variantion2}
}
\caption{Two variants of the attention layer: (a) cross-layer attention; (b) divided-layer attention.}
\label{fig:variantion}
\end{figure}

For cross-layer attention (Fig. \ref{fig:variantion1}), the scores $e_t$ and weights $\alpha_t$
are not computed using the outputs of the last LSTM layer $\{\vh_t\}_{1 \leq t \leq T}$, but the outputs of an intermediate LSTM layer \break
$\{\vh'_t\}_{1 \leq t \leq T}$,
\textit{e.g.} the second-to-last layer:
\begin{equation}
\label{eq:variation1_e}
e_t = f(\vh'_t).
\end{equation}
However, the d-vector $\ve$ is still the weighted average of the last layer output $\vh_t$.

For divided-layer attention (Fig. \ref{fig:variantion2}), we double the dimension of the last layer LSTM output $\vh_t$,
and equally divide its dimension into two parts: part-a $\vh_t^a$, and part-b $\vh_t^b$.
We use part-a to build the d-vector, while using part-b to learn the scores:
\begin{equation}
\label{eq:variation2_e}
e_t = f(\vh_t^b),
\end{equation}
\begin{equation}
\label{eq:variation2_w}
\ve = \sum_{t=1}^{T}\alpha_t \vh_t^a .
\end{equation}

\subsection{Weights pooling}
\label{sec:pooling}

\begin{figure}
  \centering
    \includegraphics[width=0.49\textwidth]{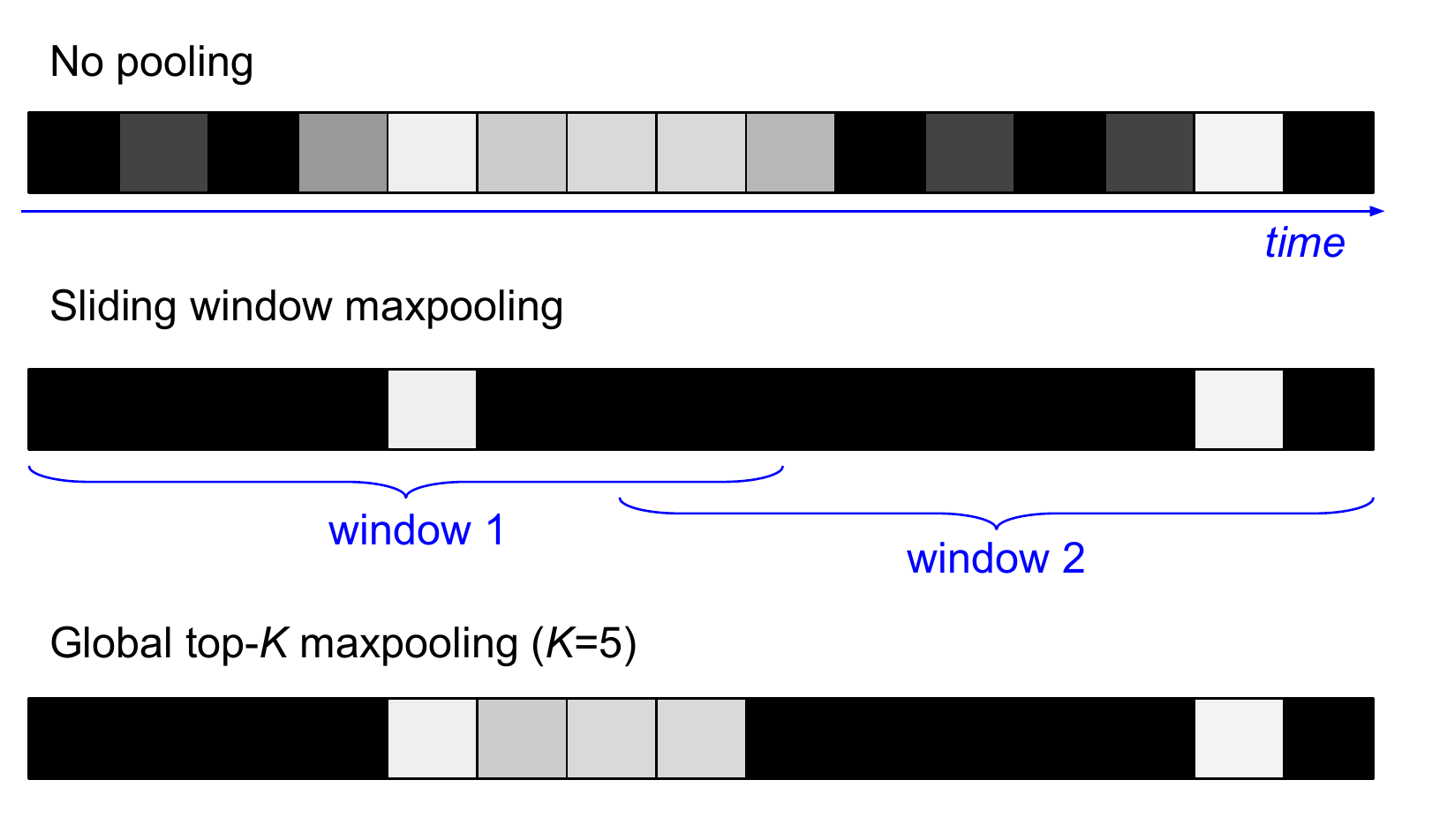}
  \caption{Different pooling methods on attention weights.
  The $t$th pixel corresponds to the weight $\alpha_t$, and
  a brighter intensity means a larger value of the weight.}
  \label{fig:pooling}
\end{figure}

Another variation of the basic attention layer is that, instead of directly using the normalized
weights $\alpha_t$ to average LSTM outputs, we can optionally perform maxpooling on the attention weights.
This additional pooling mechanism can potentially make our network more robust to temporal variations of the input signals. 
We have experimented with two maxpooling methods (Fig. \ref{fig:pooling}):
\begin{itemize}[itemsep=1mm,topsep=1mm]
\item Sliding window maxpooling: We run a sliding window on the weights, and for each window,
only keep the largest value, and set other values to 0.
\item Global top-$K$ maxpooling: Only keep the largest $K$ values in the weights, and set all other values to 0.
\end{itemize}

\section{Experiments}
\label{sec:exp}

\subsection{Datasets and basic setup}

To fairly compare different attention techniques, we use the same
training and testing datasets for all our
experiments.

Our training dataset is a collection of anonymized user voice queries,
which is a mixture of ``OK Google" and ``Hey Google".
It has around 150M utterances from around 630K speakers.
Our testing dataset is a manual collection consisting
of 665 speakers. It's divided into two enrollment sets and two verification sets for each of ``OK Google"
and ``Hey Google". Each enrollment and evaluation dataset contains respectively, an average of 4.5 and 10 evaluation utterances per speaker.

We report the speaker verification Equal Error Rate (EER)
on the four combinations of enrollment set and verification set.

Our baseline model is a 3-layer LSTM, where each layer has dimension 128, with
a projection layer \cite{sak2014long} of dimension 64.
On top of the LSTM is a linear layer of dimension 64. The acoustic parametrization consists of 40-dimensional log-mel-filterbank coefficients computed over a window of 25ms with 15ms of overlap.
The same acoustic features are used for both keyword detection \cite{prabhavalkar2015automatic} and speaker verification.

The keyword spotting system isolates segments of length $T=80$ frames (800ms)
that only contain the global password, and these segments form the tuples mentioned above.
The two keywords are mixed together using the MultiReader technique introduced in \cite{ge2e}.

\subsection{Basic attention layer}
First, we compare the baseline model with basic attention layer (Sec. \ref{sec:basic})
using different scoring function (Sec. \ref{sec:scoring}). The results are shown in
Table \ref{tab:basic}. As we can see, while bias-only and linear attention
bring little improvement to the EER, non-linear
attention\footnote{For the intermediate dimension of non-linear scoring functions, we use $m'=64$, such that
$\vW_t$ and $\vW$ are square matrices.}
improves the performance
significantly, especially with shared parameters.

\begin{table*}
\centering
  \caption{Evaluation EER(\%): Non-attention baseline model vs.
  basic attention layer using different scoring functions.}
  \label{tab:basic}
  \begin{tabular}{| c | c | c c c c c |}
    \hline
    Test data & Non-attention & \multicolumn{5}{|c|}{Basic attention}\\
    Enroll $\rightarrow$ Verify
    & baseline & $f_\mathrm{BO}$ & $f_\mathrm{L}$ & $f_\mathrm{SL}$
    & $f_\mathrm{NL}$ & $f_\mathrm{SNL}$ \\ \hline
    OK Google $\rightarrow$ OK Google   & 0.88 & 0.85 & 0.81 & 0.8  & 0.79 & 0.78 \\
    OK Google $\rightarrow$ Hey Google  & 2.77 & 2.97 & 2.74 & 2.75 & 2.69 & 2.66 \\
    Hey Google $\rightarrow$ OK Google  & 2.19 & 2.3  & 2.28 & 2.23 & 2.14 & 2.08 \\
    Hey Google $\rightarrow$ Hey Google & 1.05 & 1.04 & 1.03 & 1.03 & 1.00 & 1.01 \\
    Average                             & 1.72 & 1.79 & 1.72 & 1.70 & 1.66 & 1.63 \\
    \hline
  \end{tabular}
\end{table*}

\subsection{Variants}
To compare the basic attention layer with the two variants (Sec. \ref{sec:variant}), we use the same
scoring function that performs the best in the previous experiment: the shared-parameter
non-linear scoring function $f_\mathrm{SNL}$.
From the results in Table \ref{tab:variant}, we can see that divided-layer
attention performs slightly better than basic attention and cross-layer
attention\footnote{In our experiments, for cross-layer attention,
scores are learned from the second-to-last layer.},
at the cost that the dimension of last LSTM layer is doubled.

\begin{table}
\centering
  \caption{Evaluation EER(\%): Basic attention layer vs. variants --- all using $f_\mathrm{SNL}$ as scoring function.}
  \label{tab:variant}
  \begin{tabular}{| c | c c c |}
    \hline
    Test data & Basic $f_\mathrm{SNL}$
    & Cross-layer & Divided-layer
    \\ \hline
    OK $\rightarrow$ OK    & 0.78 & 0.81  & 0.75 \\
    OK $\rightarrow$ Hey   & 2.66 & 2.61 & 2.44 \\
    Hey $\rightarrow$ OK   & 2.08 & 2.03 & 2.07 \\
    Hey $\rightarrow$ Hey  & 1.01 & 0.97 & 0.99 \\
    Average                & 1.63 & 1.61 & 1.56\\
    \hline
  \end{tabular}
\end{table}

\begin{table}
\centering
  \caption{Evaluation EER(\%): Different pooling methods for attention weights --- all using $f_\mathrm{SNL}$ and divided-layer.}
  \label{tab:pooling}
  \begin{tabular}{| c | c c c |}
    \hline
    Test data & No pooling & Sliding window & Top-$K$ \\
    \hline
    OK $\rightarrow$ OK    & 0.75 & 0.72 & 0.72 \\
    OK $\rightarrow$ Hey   & 2.44 & 2.37 & 2.63 \\
    Hey $\rightarrow$ OK   & 2.07 & 1.88 & 1.99 \\
    Hey $\rightarrow$ Hey  & 0.99 & 0.95 & 0.94 \\
    Average                & 1.56 & 1.48 & 1.57 \\
    \hline
  \end{tabular}
\end{table}

\subsection{Weights pooling}
To compare different pooling methods on the attention weights as introduced in
Sec. \ref{sec:pooling}, we use the divided-layer attention with
shared-parameter non-linear scoring function. For sliding window maxpooling,
we experimented with different window sizes and steps, and found that
a window size of 10 frames and a step of 5 frames perform the best in our evaluations.
Also, for global top-$K$ maxpooling, we found that the performance is the best
when $K=5$. The results are shown in Table \ref{tab:pooling}. We can see that
sliding window maxpooling further improves the EER.

\begin{figure}
  \centering
    \includegraphics[width=0.49\textwidth]{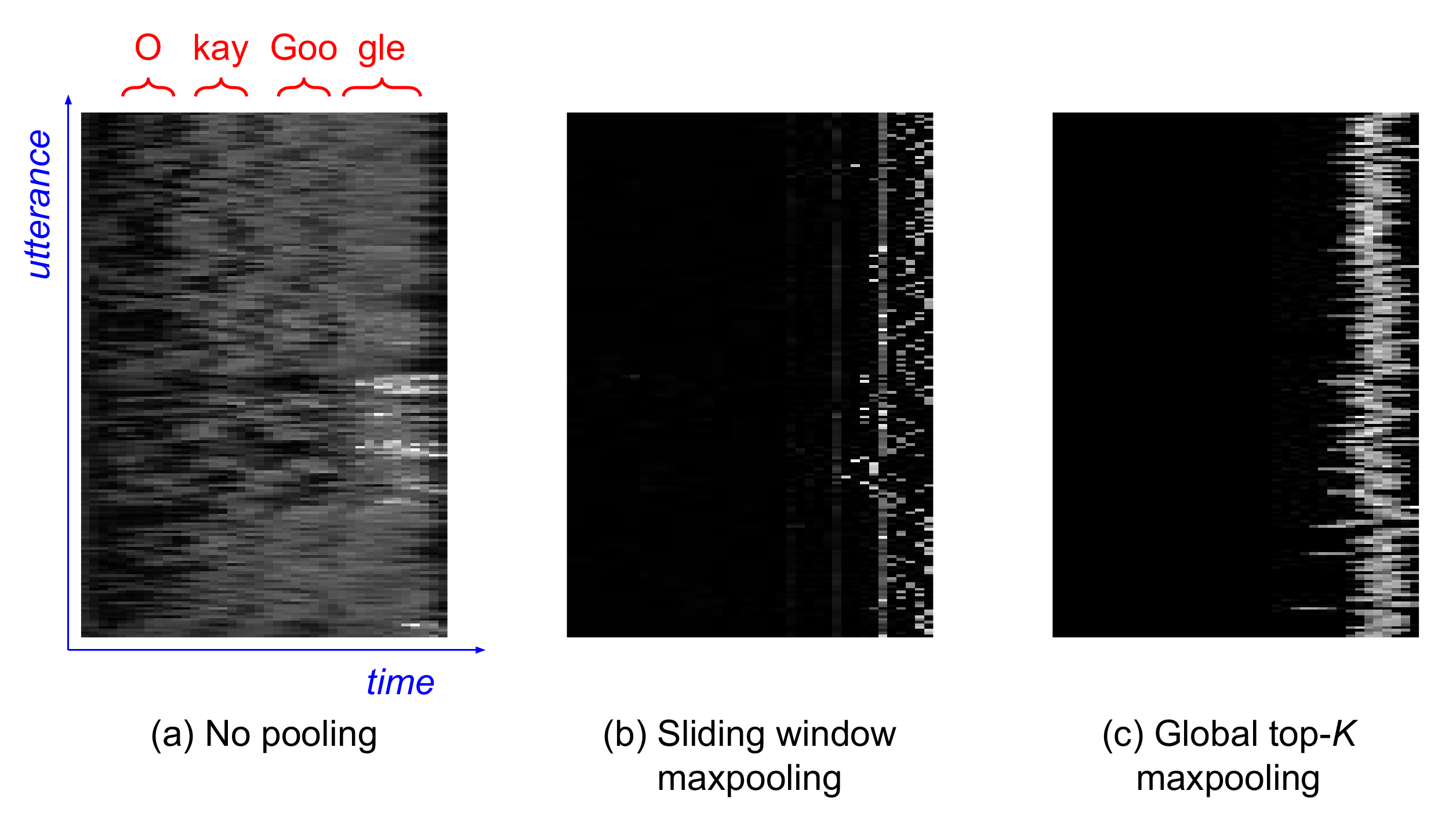}
  \caption{Visualized attention weights for different pooling methods.
  In each image, x-axis is time, and y-axis is for different utterances
  in a training batch.
  (a) No pooling; (b) Sliding window maxpooling, where window size is 10,
  and step is 5; (c) Global top-$K$ maxpooling, where $K=5$.}
  \label{fig:weights}
\end{figure}

We also visualize the attention weights of a training batch for different pooling
methods in Fig. \ref{fig:weights}. An interesting observation is that, when
there's no pooling, we can see a clear 4-strand or 3-strand pattern in the batch.
This pattern corresponds to the ``O-kay-Goo-gle" 4-phoneme or ``Hey-Goo-gle" 3-phoneme
structure of the keywords.

When we apply sliding window maxpooling or global
top-$K$ maxpooling, the attention weights are much larger at the near-end
of the utterance, which is easy to understand --- the LSTM has accumulated more
information at the near-end than at the beginning, thus is more confident to
produce the d-vector.

\section{Conclusions}
\label{sec:conclusions}

In this paper, we experimented with different attention mechanisms for our
keyword-based text-dependent speaker verification system \cite{heigold2016end}. From our experimental
results, the best practice is to: (1) Use a shared-parameter non-linear scoring function;
(2) Use a divided-layer attention connection to the last layer output of the LSTM;
and (3) Apply a sliding window maxpooling on the attention weights.
After combining all these best practices, we improved the EER of
our baseline LSTM model from 1.72\% to 1.48\%, which
is a 14\% relative improvement. The same attention mechanisms, especially
the ones using shared-parameter scoring functions, could potentially
be used to improve text-independent speaker verification models \cite{ge2e} and speaker diarization systems \cite{lstm_diarization}.

\newpage
\bibliographystyle{IEEEbib}
\bibliography{refs}

\end{document}